# Project Risk Management from the bottom-up: Activity Risk Index

**Acebes F[1], Pajares J[2], González-Varona JM[3], López-Paredes A[4]**

*Abstract*

Project managers need to manage risks throughout the project lifecycle and, thus, need to know how changes in activity durations influence project duration and risk. We propose a new indicator (the Activity Risk Index, ARI) that measures the contribution of each activity to the total project risk while it is underway. In particular, the indicator informs us about what activities contribute the most to the project's uncertainty so that project managers can pay closer attention to the performance of these activities.

The main difference between our indicator and other activity sensitivity metrics in the literature (e.g. cruciality, criticality, significance, or schedule sensitivity indices) is that our indicator is based on the Schedule Risk Baseline concept instead of on cost or schedule baselines. The new metric not only provides information at the beginning of the project, but also while it is underway. Furthermore, the ARI is the only one to offer a normalized result: if we add its value for each activity, the total sum is 100%.

[1]Fernando Acebes Senovilla (Corresponding autor)
✉ e-mail: Fernando.acebes@uva.es
GIR INSISOC. Dpto. de Organización de Empresas y CIM. Escuela de Ingenierías Industriales. Universidad de Valladolid. Paseo del Cauce 59, 47011 Valladolid (Spain). +34 983 185 833
ORCID: 0000-0002-4525-2610

[2]Javier Pajares Gutiérrez
✉ e-mail: pajares@eii.uva.es
GIR INSISOC. Dpto. de Organización de Empresas y CIM. Escuela de Ingenierías Industriales. Universidad de Valladolid. Paseo del Cauce 59, 47011 Valladolid (Spain).
ORCID: 0000-0002-4748-2946

[3]José Manuel González Varona
✉ e-mail: josemanuel.gonzalez.varona@uva.es
GIR INSISOC. Dpto. de Organización de Empresas y CIM. Escuela de Ingenierías Industriales. Universidad de Valladolid. Paseo del Cauce 59, 47011 Valladolid (Spain).

[4]Adolfo López Paredes
✉ e-mail: aparedes@eii.uva.es
GIR INSISOC. Dpto. de Organización de Empresas y CIM. Escuela de Ingenierías Industriales. Universidad de Valladolid. Paseo del Cauce 59, 47011 Valladolid (Spain).
ORCID: 0000-0001-5748-8308

## 1. Introduction

Project Management is the application of knowledge, skills, tools and techniques to forecast activities to meet project requirements that is accomplished through the proper application and integration of different project managements (Project Management Institute, 2017). An important group of processes that allows organizations to execute projects effectively and efficiently includes those related to project monitoring and control. They include the process of tracking and reviewing the project's progress to satisfy the project management plan requirements and to achieve stakeholder satisfaction.

Continuous monitoring allows the project management team to know the project's health and to identify those areas that may require special attention. Control includes determining preventive or corrective actions, or modifying action plans and their follow-up, to establish if the performed actions allowed the problem to be solved.

Over the years, the Earned Value Methodology (EVM) has been widely used to control projects in both costs and time terms. This methodology is based on measuring the deviation of the current time or the ongoing project's cost from the planned value baseline (for an overview, see e.g. Anbari (2003), Fleming and Koppelman (1998) or Pajares and López-Paredes (2011)). In many cases, however, project managers may be interested in identifying the critical programming components that could have the strongest impact on project objectives (Vanhoucke, 2012a). To carry out this process, the SRA framework is often applied.

If Project Risk Management includes the processes to carry out the management planning, identification, analysis, response planning, response implementation and monitoring of project risks (Project Management Institute, 2009), the SRA relates information on the risks (uncertainty) of project activities to baseline scheduling, and provides information on the sensitivity of individual project activities so that the potential impact of the uncertainty of activity on the final project duration can be assessed. It identifies the key components to successfully complete the project on time by providing this information before the project has even started. Identifying the activities that are most sensitive to the project allows project managers to focus primarily on those that matter (Vanhoucke, 2012b). This will enable a more accurate response during project monitoring or control, and should contribute positively to overall project performance.

The SRA is a project management simulation technique for assessing the uncertainty of schedule compliance that helps to forecast the impact of time and cost deviations on project objectives (Vanhoucke, 2015). The SRA relates the information obtained on the risk of project activities to the planned timeline, and facilitates information on the sensitivity of project activities being obtained in such a way that it can be useful for assessing whether the uncertainty of activities may impact the final project duration. This information is useful for project managers because they can use it to determine the level of attention that they should pay to an activity given its influence on the final project duration (Vanhoucke, 2016).

Hulett (1996) describes the sequence of activities that must be followed to implement the SRA procedure:

— The CPM Schedule. Defining the scheduling baseline is the key component of quantified risk assessment because it acts as a reference point for all the calculations made during the subsequent simulation. It provides information on the expected project duration, the start and end dates of activities, and on the use of several types of resources over time

— Determining the uncertainty of activities. The duration of activities is subject to a margin of error (uncertainty), which leads to unexpected variations in the duration of activities. For this purpose, the most appropriate distribution function of the activity's behavior is determined, which will be incorporated into the simulation phase

— Simulating project planning. Once the distribution functions of the duration corresponding to each activity have been determined, Monte Carlo Simulation is applied as one of the most widespread probabilistic techniques for conceptual forecasts of durations and decision making (Chou, 2011; Liu and Wang, 2013). During each simulation, a random duration is assigned to each activity according to its distribution function. Therefore for each simulation, project duration differs

— Sensitivity analysis. During each simulation, duration data are collected for all the project activities and total project duration. With these data, a sensitivity analysis is carried out to know the influence of each activity on total project duration in relation to each activity belonging to the project's critical path or not. The result obtained for each activity informs about the importance of this activity for the project and how variation in the parameters of the former affects the latter

Project managers need to be able to discriminate the activities that are the most influential for the whole project as they will contribute more to the possible final variation in any of the scheduling objectives (Vanhoucke, 2015). In addition, this should serve as a basis for proactive decision making during project monitoring. The ultimate objective of these studies is to provide project managers with valuable information to allow them to know which activities are the most influential and determinant for project development in each case in order to take the appropriate preventive measures.

The SRA framework was extended by Vanhoucke (2010b, 2012a, 2012b, 2016) to a broader process called Dynamic Scheduling, which includes Project Control. His proposal is founded on three stages:

— Baseline Scheduling: it implies developing a schedule using information from project activities (start and end date of each one) by considering each activity's duration, its precedence ratio, the resources available to perform activities, as well as other project characteristics, to find an appropriate project schedule
— Risk analysis: it consists of obtaining information on the sensitivity of activities in respect to the total project duration by considering the uncertainty of the activities themselves. This analysis makes it possible to examine the impact that variations in the duration of activities would have on the project objectives
— Project Control: it allows us the progress of the project schedule to be checked. This monitoring is carried out using the information obtained in previous steps and should be useful for performing corrective actions if problems arise

From Vanhoucke's point of view (2012b), the usefulness of basic project scheduling is quite limited and only acts as a reference point in the project's life cycle. Therefore, a project schedule should only be considered to be a predictive model that can be used for time and cost risk analyses, project control and project performance measurement. Thus Dynamic Scheduling brings together scheduling, risk analyses and project control in a single methodology.

In this paper, we propose a new metric to be used with both the SRA and Dynamic Scheduling frameworks for risk analyses. This metric is built on the Project Risk Baseline (Pajares and López-Paredes, 2011) instead of on a Project Schedule Baseline (see Fig. 1), and allows a different approach to estimate risk analyses from the bottom-up. In our case, this indicator will be calculated according to the risk with which each activity contributes to the project's total risk.

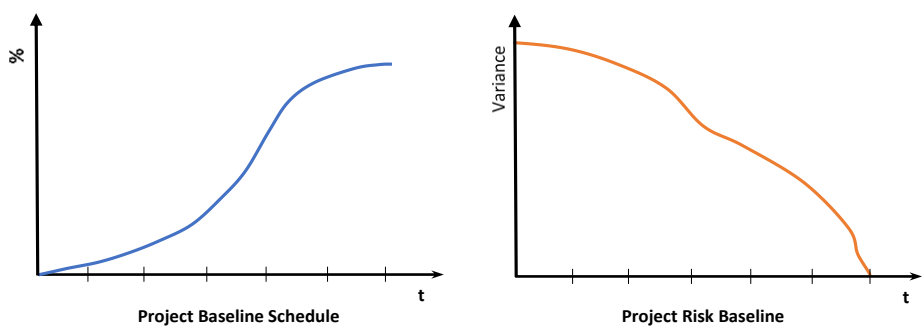

**Fig. 1** Representation of the Project Baseline Schedule vs. the Project Risk Baseline.

The proposed index, as it will be calculated, will consider the activity's risk (variability). Unlike other measures that take into account only the project's risk (uncertainty) at its initiation time, this indicator measures the evolution of the project's risk level according to its execution, but will also take into account the project network characteristics and the situation of the activity in that network. This aspect, which was initially observed by Tavares et al. (2002), was also subsequently considered by Madadi and Iranmanesh (2012) for the calculation of the MOI index that they proposed.

The paper is organized as follows. First we review the most relevant metrics used in the project risk analysis. We briefly review their advantages and disadvantages, as well as the mathematical notation used to calculate them. In the next section, we present our contribution: a new sensitivity metric called the Activity Risk Index (ARI). The following section presents a discussion of results and provides significant examples that we use to provide readers with a better understanding of the calculation and results offered by this indicator. Finally, we present the conclusions drawn from our research.

## 2. Literature review

The literature on activities' sensitivity metrics is broad and diverse. Since the well-known PERT methodology was proposed in the late 1950s, research on assessing the sensitivity of project activities has received increasing attention (Vanhoucke, 2011). Furthermore, forecasting project duration has become a critical issue for project managers as the traditional Critical Path Analysis provides unrealistically optimistic results that are not consistent with the probabilistic results of simulations (Klingel, 1966; Schonberger, 1981; etc.).

The analysis of the influence of activities on project results is not new. Some authors made an initial classification of studies on deterministic networks and studies on stochastic networks (Madadi and Iranmanesh, 2012). The proposal of the PERT methodology and the development of the CPM model can be considered the first solutions to use deterministic networks. In the CPM model, activities that are on the critical path are considered critical activities. Later studies analyzed the criticality of project activities and the whole project mainly by fuzzy methods (Jassbi et al., 2008; Kuchta, 2001).

The most important studies focus on using stochastic networks. Martin (1965) defines an activity's criticality as the probability of the activity belonging to the longest path. To do this, he proposes the Activity Criticality Index (ACI) and previously used the Path Criticality Index (PCI) concept to identify the probability of a path having the longest duration.

The indicators proposed by Martin (1965) generate many subsequent studies, largely due to difficulty and mathematical demand when making calculations to determine each critical path and, consequently, critical activities (Bowman and Muckstadt, 1993; Chanas and Zieliński, 2002; Ghomi and Teimouri, 2002; Van Slyke, 1963, etc). For example, Dodin and Elmaghraby (1985) make a rough calculation of the ACI without having to previously calculate the PCI. Other authors follow various techniques, such as Monte Carlo simulation with mathematical analysis, to estimate an activity's criticality (Bowman, 1995). Some others use fuzzy techniques to calculate the critical path by applying statistical tools with which to prioritize project activities (Lin and Yao, 2003) by employing linear programming (Chen, 2007) or analytical and fuzzy methods, together with the PERT methodology (Chen and Huang, 2007).

Despite the importance and usefulness of the indicators proposed by Martin (1965), the resultant values are not always consistent with the actual project completion probability (Liu and Wang, 2013). For this reason, Williams (1992) proposes two new indicators: the Cruciality Index (CrI) and the Significance Index (SI). With them, he intends to prioritize the importance of activities by relating uncertainty in completing the project to the coefficient of the linear correlation between each activity's duration and project duration. However, the proposed CrI indicator can only describe the linear relation between each activity's duration and project duration. Cho and Yum (1997) observed this nonlinear relation between the two variables. Subsequent studies found that this indicator only considered the effect of the activity's variability on project duration (Elmaghraby, 2000). An extreme case is that in which a deterministic activity belongs to the critical path, but the value of its CrI will always be zero.

Other studies on the criticality of activities in PERT networks appear in the literature. Cui et al. (2007) propose the Activity Critical Comprehensive Index (ACCI) to assess each activity's criticality from three perspectives: the activity's duration, variance, criticality. Finally, Madadi and Iranmanesh (2012) propose the Management Oriented Index (MOI) to measure the importance of activities by incorporating criteria of the variability (risk) of activities, the effect of activities on average project duration and the morphological characteristic of the project network.

### 3. Activity Sensitivity Metrics for Risk Analyses

We explain the metrics that are broadly considered within the SRA framework for project risk analyses, calculated by applying Monte Carlo Simulation: Criticality Index (CI), Cruciality Index (CrI), Significance Index (SI); Schedule Sensitivity Analysis (SSI). We consider another metric that extends the SSI to contemplate the activities network topology: the Management-Oriented Index (MOI).

### 3.1 Criticality Index (*CI*).

With this indicator, we measure the influence of an activity's duration on total project duration. It is calculated as the percentage of the simulations with which this activity belongs to the critical path. It was introduced by Martin (1965) and frequently appears in the literature:

$$CI = P(tf_i = 0) \tag{1}$$

where *CI* is the Criticality Index and *tf$_i$* is the total float of activity *i* (null value for its slack).

This sensitivity indicator has been widely used over the years, but is not significant itself as Williams (1992) demonstrated. The main disadvantage of the Criticality Index is that its measurement offers us a probability value and leaves aside the other dimension of the activity; that is, the impact that this activity may have on total project duration, which is why this information should be complemented with other metrics.

**3.2 Cruciality Index (CrI)**

Williams (1992) focused on measuring the importance of the activity calculated as the correlation between the activity's duration and project duration.

$$rI = |corr(d_i, PD)| \tag{2}$$

If project duration extends when the duration of an activity is longer and project duration is shorter when the activity ends in less time, the Cruciality Index will be high.

The three types of cruciality indices are important:

*a) CrI - Pearson product-moment correlation*

This is the most widely used Cruciality Index, although it may be advisable to use other more appropriate indicators due to the nonlinearity between an activity's duration and project duration (Cho and Yum, 1997).

$$CrI(r) = \frac{cov(d_i, PD)}{var(d_i) \cdot var(PD)} \tag{3}$$

where *cov(x,y)* is the covariance between values *x* and *y*, and *var(x)* is the variance of value *x*.

*b) CrI - Spearman's rank correlation*

In an attempt to correct the above problem, the indicator calculated according to this formula takes into account possible nonlinearities by converting the values of variables into ranges.

$$CrI(\rho) = E\left(1 - \frac{6\sum_{k=1}^{nrs} \delta_k^2}{nrs(nrs^2-1)}\right) \tag{4}$$

where *nrs* is the number of Monte Carlo simulations, $\delta_k$ is the difference between the classification values of *d$_i$* and *PD* during simulation k.

*c) CrI - Kendall's tau rank correlation*

It measures the degree of correspondence between two rankings and evaluates the importance of this correspondence.

$$CrI(\tau) = \frac{4P}{nrs(nrs-1)} - 1 \qquad (5)$$

where *P* is used to represent the number of matching pairs of activity duration and project duration.

### 3.3 Significance Index (SI)

This indicator (Williams, 1992) intends to show the importance of individual activities for overall project duration. It incorporates an estimation of the potential impact that a delay in activity can have on the whole project.

$$SI = E\left(\frac{d_i}{d_i + tf_i} \cdot \frac{PD}{E(PD)}\right) \qquad (6)$$

where *E(x)* is used to identify the expected value of *x* and $d_i$ corresponding to the duration of activity *i* in the simulation. *PD* is the project duration in the simulation.

Although this indicator provides more relevant information on the importance of activities in relation to the total project, it is not considered the definitive index.

### 3.4 Schedule Sensitivity Index (SSI)

This indicator (Project Management Institute, 2004) relates the probability of an activity belonging to the critical path (i.e. probability) corrected with the relation between the variability of duration and that of the project (i.e. the impact of the activity on project duration).

$$SSI = CI \cdot \frac{\sigma_{d_i}}{\sigma_{PD}} \qquad (7)$$

A study by Vanhoucke (2010b) worked on the above-explained metrics by comparing their effectiveness in monitoring simulated projects during their execution. It concluded that the SSI indicator is the most realistic when prioritizing activities based on their sensitivity to project duration.

### 3.5 Management-Oriented Index (MOI)

Although we focus on the metrics used in the SRA and Dynamic Scheduling, there are other sensitivity metrics in the literature about project monitoring: Cho and Yum (1997); Elmaghraby et al. (1999); Gutierrez and Paul (2000); Kuchta (2001); Tavares et al. (2004); Cui et al. (2007). Among them, we considered including in the benchmark a new sensitivity metric that incorporates information on the project network structure (Madadi and Iranmanesh, 2012).

$$MOI_i = \frac{\sigma_i}{\sigma_{max}} \frac{1}{(E(TF_i) - Post\_Density_i + 1)} \qquad (8)$$

where *E(TF$_i$)* is the expected slack value of activity *i*. *Post_Density$_i$* equals the total number of successors of activity *i* divided by the total number of project activities.

| Metric | Mathematical expression | Benefits and criticalities |
|---|---|---|
| Criticality Index (CI) | $CI = P(tf_i = 0)$ | The criticality index measures the probability that an activity is on the critical path. It measures the influence of this activity's duration on the total project duration. Attempts should be made to short those activities with a high CI as they are likely to become bottlenecks. The main drawback of the CI is that its focus is limited to measuring probability, which does not necessarily mean that the activities with a high CI value strongly impact overall project duration |
| Cruciality Index (CrI) | $CrI = |corr(d_i, PD)|$ | Measures the correlation between an activity's duration and total project duration. It represents the relative importance of an activity and estimates the uncertainty of total project duration being due to the uncertainty of an activity. |
| CrI - Pearson product-moment correlation | $CrI(r) = \dfrac{cov(d_i, PD)}{var(d_i) \cdot var(PD)}$ | |
| CrI - Spearman's rank correlation | $CrI(\rho) = E\left(1 - \dfrac{6\sum_{k=1}^{nrs} \delta_k^2}{nrs(nrs^2 - 1)}\right)$ | Pearson product-moment is a correlation measure of the degree of the linear relation between two variables. However, the relation between an activity's duration and total project duration usually follows a nonlinear relation. Spearman rank correlation coefficient or Kendall's tau measure is a measure of nonlinear correlation. |
| Kendall's tau rank correlation | $CrI(\tau) = \dfrac{4P}{nrs(nrs - 1)} - 1$ | The IRC (τ) assumes that the values of variables (i.e. duration of activities and project duration) are converted into categories. It measures the degree of correspondence between two categories and assesses the importance of this correspondence. Special attention should be paid to activities with a high Cruciality Index due to the uncertainty they introduce in the project |
| Significance Index (SI) | $SI = E\left(\dfrac{d_i}{d_i + tf_i} \cdot \dfrac{PD}{E(PD)}\right)$ | The Significance Index aims to expose the importance of individual activities for total project duration, rather than expressing the criticality of an activity through the probability concept. It estimates the potential impact that a delay in activity may have on the entire project. |
| Schedule Sensitivity Index (SSI) | $SSI = CI \cdot \dfrac{\sigma_{d_i}}{\sigma_{PD}}$ | It combines the standard deviations of activity duration and project duration with the CI. The SSI provides relatively better results than other sensitivity indices when assessing the contribution of the corrective actions taken during project monitoring |
| Management-Oriented Index (MOI) | $MOI_i = \dfrac{\sigma_i}{\sigma_{max}} \dfrac{1}{(E(TF_i) - Post\_Density_i + 1)}$ | The MOI assesses the importance of activities by considering the their variability (risk), the effect of activities on average project duration and the project network's morphological characteristic. |
| Activity Risk Index (ARI) | $ARI_i = \dfrac{SRV_0 - SRV_i}{SRV_0}$ | The ARI prioritizes the importance of project activities according to the uncertainty that they contribute to the entire project by considering each period of its execution, from the beginning to the end. To calculate this, not only is each activity's uncertainty taken into account, but also the network structure and the location of each activity within the network. |

**Table. 1** Summary of the most popular methods for prioritizing project activities.

The indicator was used by the authors to conduct a comparison study along with other existing indicators. They concluded that this proposed indicator provides the most representative metric for the sensitivity analysis of activities.

Table 1 summarizes the methods for prioritizing activities most widely used by researchers, and for this work in the case study section. The mathematical expression of all the indicators and their main characteristics are shown in this table.

### 4. Project Risk Analysis from the bottom-up: Activity Risk Index (ARI)

The SRA framework takes the Project Baseline Schedule as the starting point. Similarly, the Project Cost Baseline would be considered, although metrics are defined in "time" terms (project duration).

We developed an alternative considering the Project Risk Baseline as Cagno et al. (2008), Pajares and López-Paredes (2011), Acebes *et al.* (2013, 2014b) did. We are interested in computing the contribution of activities to the total project risk while the project is underway.

**4.1 Schedule Risk Baseline (SRB)**

The Risk Baseline represents the evolution of the project's risk value throughout its life cycle. That is, the risk (uncertainty) to comply with the other project activities.

The project's risk at a given time (Actual Time – AT) is calculated as the uncertainty (measure as a variance) provided by the activities not yet completed (from the AT instant to the end of the project) by taking into account that a project's efficiency is calculated according to its planning, and until the time it is considered (see Figure 2).

In this figure, we represent a project with four activities (A1, .., A4), and each activity's duration and the probability distribution of their duration (Fig. 2a). With this data and the cost of each activity, we can compute the project cost baseline (or Planned Value, PV). Thus we can compute the Budget at Completion (BAC) at the final scheduled time (Schedule At Completion - SAC) (Fig. 2b). We show the SRB at the bottom of the figure (Fig. 2c). AT corresponds to the project's current execution time.

In Fig. 2a we represent activity (or part of it) as an unfilled rectangle if the activity has already been executed up to instant AT. Its duration will be deterministic.

If the activity (or part of it) has not yet been executed, it is represented as a rectangle filled in with a color. This activity (or part of it) continues to confer the project uncertainty.

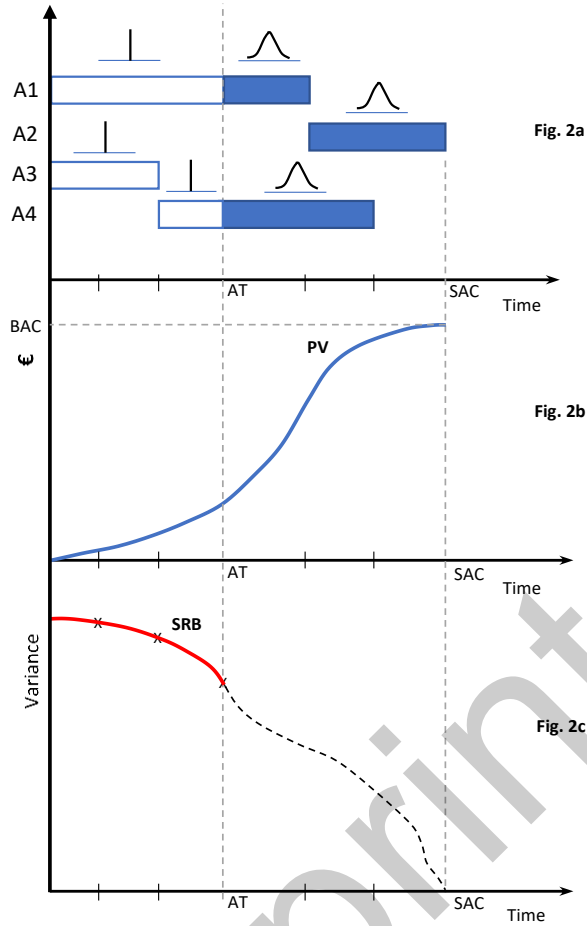

**Fig. 2** Graphical representation of the Schedule Risk Baseline calculation process.

For each execution period, we compute the project's uncertainty as the value of the variance of the output distribution function of the project schedule: Eq. 11. As we assume that project activities are subject to uncertainty during their duration, we use Monte Carlo simulation to obtain the distribution functions of total project duration. From the statistical data of these distribution functions, we extract the data of the variance that we use as a value of the project's risk during each period.

$$PD = PD(Gr, \sigma_i^2) \tag{9}$$

$$SAC = AT + PD(Gr, \sigma_i^2)\big|_{t=AT}^{t=SAC} \tag{10}$$

$$SRB_{AT} = var\left(PD(Gr, \sigma_i^2)\big|_{t=AT}^{t=SAC}\right) \tag{11}$$

where *PD* is the project total duration function, which depends on the project network (*Gr*) and the variance of project activities; *SAC* (Schedule At Completion) is the planned project duration and *SRB*$_{AT}$ is the value of the SRB at instant AT.

We consider that running the project between each time period is carried out according to the initial planning (as defined for any baseline). After performing Monte Carlo simulation during each project execution period, we obtain the probability distribution functions of total project duration. Using the statistics of these distribution functions, we extract the variance data from these graphs. These variance data during periods are those we transfer to the representation of the baseline of programming risks.

During each time period, the project's variance is calculated by applying Monte Carlo simulation. At this point, the activities that confer the project uncertainty are those that have not yet been executed (represented in Fig. 2a as colored rectangles).

As the project progresses, some activities finish and their duration becomes deterministic (Activity A3 and the uncolored area of activities A1 and A4 in Fig. 2a). The remaining activities still pending will continue to confer the project uncertainty (Activity A2 and the colored area of activities A1 and A4 in Fig. 2a), which will be reflected in the new value of variance during its corresponding execution period. When the project ends (t=SAC) and all the activities have been executed, uncertainty will be null. Therefore, the value of variance will also be null.

Finally by joining all the points corresponding to the $SRB_{AT}$ values at each time instant (AT), we construct the SRB graph (represented in Fig. 2c).

**4.2 Activity Risk Index (ARI)**

The ARI is an indicator that measures the risk (uncertainty) with which each activity contributes to the entire project. Once again, risk is measured as the variance of project duration during each execution period. To calculate this indicator, we use the Risk Baseline concept that we explained in the previous section.

We summarize the methodology used by the authors to obtain the Activity Risk Index (ARI) for each project activity. We started by employing the information provided from each project activity (duration, cost, sequencing, uncertainty) to represent the SRB. From Fig. 2c, we assume that we represent the SRB of the planned project ($SRB_0$). This graph provides information on the project's risk provided by the activities pending execution at each control time.

We calculate "Total_Risk" as the sum of the variance of the project duration during each period, from the beginning of the project until it ends, calculated in the planning phase. We represent it by the Schedule Risk Value concept.

We define the Schedule Risk Value (SRV$_0$) as the area under the *SRB$_0$* curve, where this curve is the SRB of the planned project when all the activities contribute to the project with the initial uncertainty with which they were programmed. Equation 12:

$$SRV_0 = \int_{t=0}^{t=SAC} SRB_0 \qquad (12)$$

where SRV$_0$ is the "total risk" of the planned project, SRB$_0$ is the curve that represents the elimination of risk (uncertainty) from the planned project, and SAC is the planned project duration (Schedule At Completion).

We must calculate the risk with which a particular activity contributes to the total project's risk (SRV$_i$). To this end, we recalculate the Risk Baseline, but by considering that the duration of this activity is deterministic. By considering the deterministic duration activity, we eliminate the uncertainty that this activity adds to the entire project.

In other words, activity does not confer the project uncertainty in this new simulation. In Fig. 2c we represent the SRB$_i$ curve. This curve is the result of calculating the project's Risk Baseline by considering, for example, that project activity *"i"* is deterministic. We recalculate the total risk (SRV$_i$) as the area under the SRB$_i$ curve.

$$SRV_i = \int_{t=0}^{t=SAC} SRB_i \qquad (13)$$

In Fig. 3 we offer an example to explain the meaning of the SRV concept. The area under the SRB$_0$ curve, identified by blue stripes, is the total risk of the planned project (SRV$_0$). The area under the SRB$_i$ curve, drawn with red stripes, represents the project's risk when considering activity "i" of deterministic duration (SRV$_i$).

The risk contributed by activity "i" to the project results from the subtraction of the two previous values, and is the equivalent to the area between both curves represented in Fig. 3.

In this way, the Risk Index of activity "i" (ARI$_i$) is obtained, as seen in Eq. 14:

$$ARI_i = \frac{SRV_0 - SRV_i}{SRV_0} = \frac{\int_{t=0}^{t=SAC} SRB_0 - \int_{t=0}^{t=SAC} SRB_i}{\int_{t=0}^{t=SAC} SRB_0} \qquad (14)$$

where SRV$_0$ is the area under the *SRB$_0$* curve, SRV$_i$, is the area under the SRB$_i$ curve by considering that activity "i" is deterministic. The proposed indicator represents this magnitude expressed as a decimal and can also be represented as %.

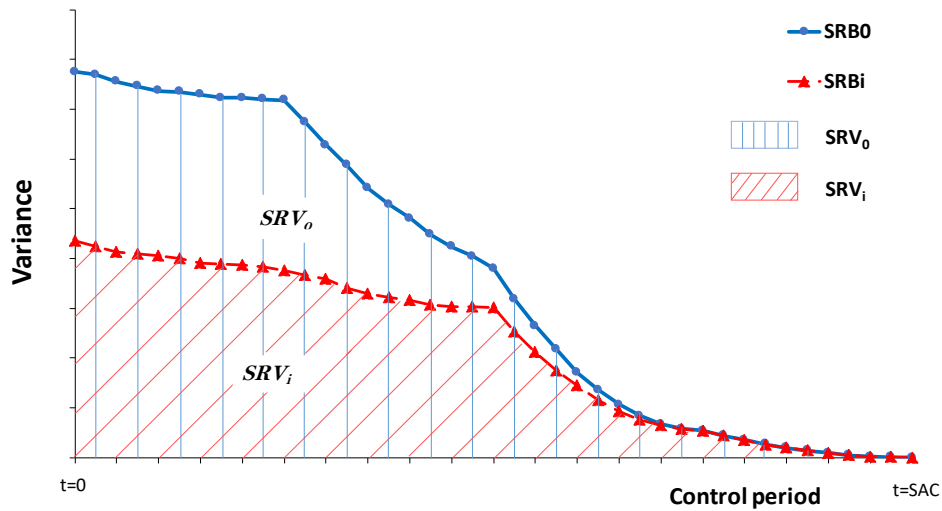

**Fig. 3** Planned Schedule Risk Baseline ($SRB_0$) and the project's Schedule Risk Baseline if activity $A_i$ is 'deterministic' ($SRB_i$)

We could repeat the operation for each project activity by considering that activities have initially planned uncertainty and by assigning a deterministic value to the new activity for which we wish to calculate its corresponding indicator. The $ARI_i$ values obtained for each activity are normalized so that the sum of them all is the unit (or 100%).

The proposed indicator (ARI) prioritizes activities by ranking them according to the uncertainty with which the activity contributes to the overall project. The value of the indicator for each activity depends on its planned uncertainty. Furthermore, the calculation of the indicator takes into account the project network structure and its position in that network (whether the activity is on a serial or parallel path). Finally, the instant at which the activity is executed also has an influence.

We use a very simple project to explain how the proposed metric (ARI) operates. The project consists of two activities in series: first A1 and then A2. Both have the same statistical properties: duration and uncertainty (time distribution function, mean, variance). For both activities, the example considers that duration follows a normal distribution function with mean 5-time units and 0.64 variance. In Fig. 4 we represent the project's SRB ($SRB_0$) and two additional curves for this project, $SRB_1$, which is calculated like SRB, but A1 is considered deterministic ($SRB_1$); $SRB_2$, when A2 is the deterministic activity.

As previously explained, we calculate the ARI for activity $A_1$ (Eq. 14) as the relation between $SRV_1$ and $SRV_0$ (($SRV_0-SRV_1$)/$SRV_0$). In the same way, the ARI for activity $A_2$ is calculated as ($SRV_0-SRV_2$)/$SRV_0$. Figure 4 graphically shows how $SRV_2$ is higher than $SRV_1$. Consequently, $ARI_{A2}$ is higher than $ARI_{A1}$. This is due to the relative position of A2 and A1 in the activities network because both activities have the same statistical properties. The difference in the ARI metric value is explained by the relative position that both occupy in the network.

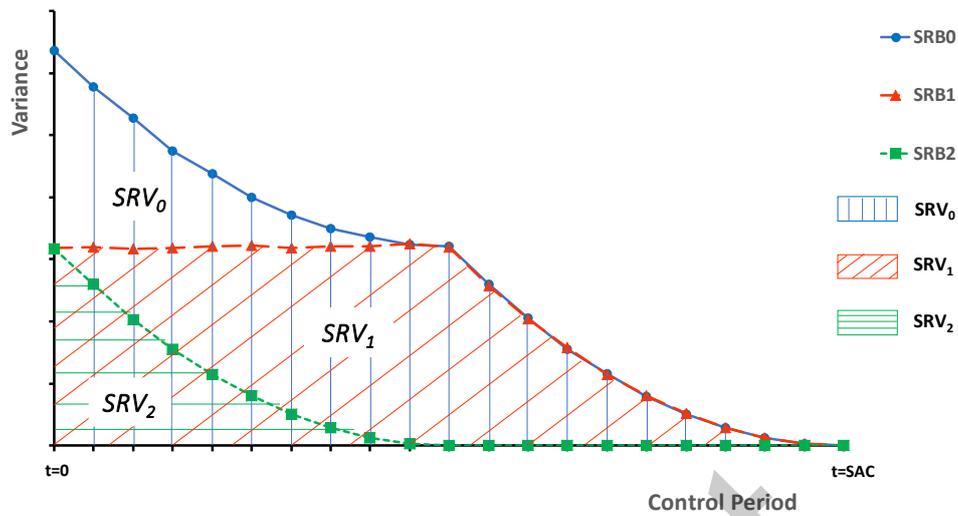

**Fig. 4** SRB and SRV for a project with serial activities

The ARI is a metric that facilitates prioritizing activities for risk planning and control in "probability x impact" terms. We use Monte Carlo simulation to capture the "probability" factor, and the SRB to measure the "impact" factor due to the statistical properties and the position of each activity in the network. It facilitates a bottom-up approach to Project Risk Management from activities.

5. **Case Study**

In this section we show the results of applying the calculation of different indicators to a project by focusing our explanation on the process followed to obtain the data required to compute the indicator we propose. As the objective of this section is to show how to calculate indicators, we choose a simple project for didactic purposes. In this case, we use the project with five activities according to Fig. 5. The initial and final activities (A0 and A6) are fictitious, and represent the start and end of the project.

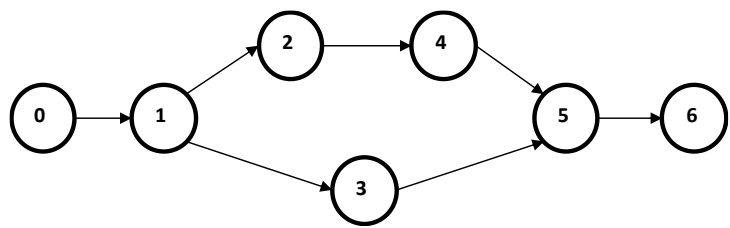

**Fig. 5** AON project diagram

The parameters with which we define each activity in our model are described in Table 2. We assume that project activities cannot be fractionated or divided, and are scheduled to begin as soon as possible based on their precedence relations.

To formalize the uncertainty type of activities, we consider aleatory uncertainty, which generates a feasible range of possible results. Any other uncertainty type (stochastic or epistemic) can be considered, and should be included in the programming of each activity. We assume that the duration of the network activities in this example is modeled as normal distributions, one of the most widely used in the project literature. This does not prevent the activity from being characterized by a different distribution function type (triangular, beta, etc.). In this case, the corresponding parameters would be included in its programming.

The main objective is to determine the value of the Activity Risk Index (ARI) indicator for each activity and to represent the intermediate graphs used to calculate it. We also incorporate the calculation of the other main sensitivity indicators and we discuss the priority of each activity according to the employed indicator.

To calculate the ARI, the first step is to calculate the Risk Baseline of the planned project ($SRB_0$). We use Monte Carlo simulation applied to the project, whose activities are programmed according to the data indicated in Table 2.

| Activity | μ (time units) | σ (time units) | Precedents Activities |
|---|---|---|---|
| A1 | 5 | 0,4000 | - |
| A2 | 5 | 0,7000 | A1 |
| A3 | 10 | 1,4000 | A1 |
| A4 | 5 | 1,2000 | A2 |
| A5 | 5 | 0,4000 | A3, A4 |

**Table 2** Characteristics of project activities

Once we obtain the representation of curve $SRB_0$ (in Fig. 6 in red), we then calculate the area under this curve, which represents the project's Schedule Risk Value ($SRV_0$); i.e. the sum of the project variances at each control time. The project is executed according to planning.

Then we calculate the $SRB_i$ for each activity. To do so, and by calculating each activity individually, we consider its duration to be constant (deterministic) and to equal most of the probable value with no variability. We represent all these curves in Fig. 6. The proposed indicator represents the percentage, in relation to the total, of the risk that eliminates each activity by removing its uncertainty; that is, by taking it to be deterministic.

If an activity is deterministic, it does not confer the whole project uncertainty. Hence we obtain a different curve than that planned. The larger the area between the two curves, the higher the risk to be eliminated.

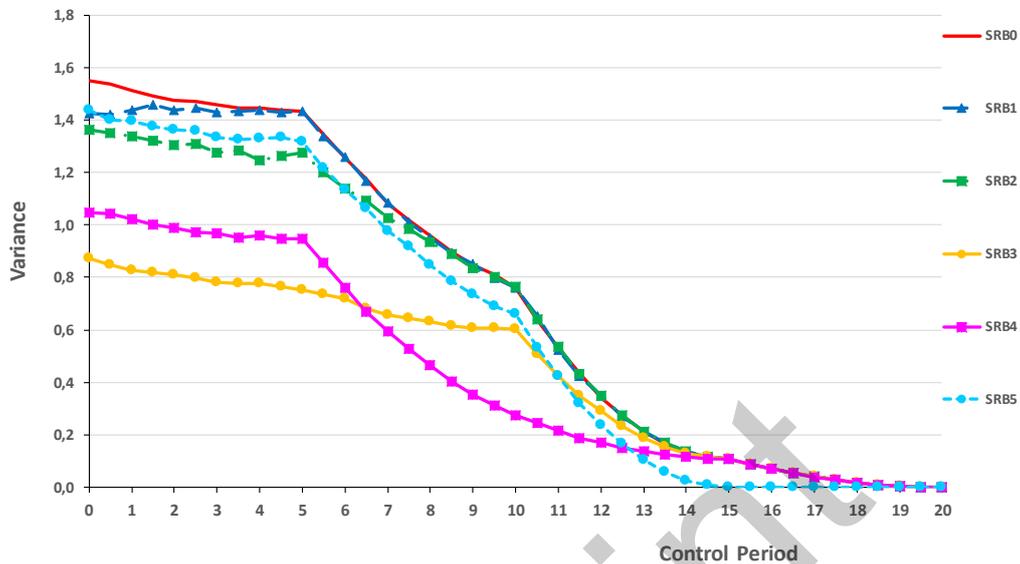

**Fig. 6** SRB representations for calculating the ARI indicator

We can focus on activity A1, at the beginning of the project network. When considering the duration of this activity to be deterministic, we first observe that the uncertainty at the beginning of the project ($SRB_1$) is less than that of the original project ($SRB_0$), because we eliminate the uncertainty that A1 confers. We also observe that the project's uncertainty remains constant until period 5, precisely when this activity ends and activities A2 and A3 start being executed. As A1 is considered deterministic, uncertainty is not eliminated while it is run, and all the uncertainty of the subsequent activities still pending execution continues to exist. From this point onward, the project's uncertainty decreases, which coincides with the original project. We see that A1 only confers the project uncertainty in the initial stretch, between instants t=0 and t=5. By calculating the difference in areas (using Eq. 14), we obtain the importance of this activity as far as the project's total risk is concerned.

If we look at the representation of activity A5 ($SRB_5$), we see that uncertainty is lower than that of the original project from the beginning of the project because activity A5 does not confer uncertainty. Additionally, the project's diminished uncertainty remains constant until instant t=15, which becomes null. This means that all the activities that confer the project uncertainty have been executed and, from instant t=15 to instant t=20 at the end of the project, uncertainty is 0. This period corresponds to the execution of activity A5, whose duration is considered deterministic. We then calculate the degree of importance of this activity for the project's risk by calculating the difference of areas between the original ($SRB_0$) and that corresponding to this activity ($SRB_5$).

The other activities are also represented in Fig. 6. Each one has a specific layout, which depends on the uncertainty that we eliminate by turning them into determinists. It also depends on the position in the network (i.e. if it is on a serial or parallel path), and on the location in the project network. Obviously, the representation of the curve is related to the uncertainty programmed for the activity. The greater the uncertainty that it confers the project, the lower the graph value at the source.

An activity that is executed during the first project moments quickly eliminates its uncertainty, which occurs with activity A1. The uncertainty conferred to the project by activity A1 is maximum at the initial instant and decreases while being executed. The uncertainty that this activity adds to the project will be zero when the execution of this activity ends. For these activities, the uncertainty that they add to the project remains constant until their execution begins; e.g., activity A5.

Uncertainty begins to diminish from the time when the execution of an activity begins. Once the execution of the activity ends, it no longer adds uncertainty to the project. However, we must bear in mind that the activity (e.g. A5) has added uncertainty to the project from the time the project started until the activity ended.

This implies that two activities with the same characteristics in duration and uncertainty terms (e.g., A1 and A5), and the fact that they occupy different positions in the network, influence the importance that we should attach to each one,

Having explained how we obtained the graphs in Fig. 6, we then calculated the ARI metric for each activity. To do so, we calculated the area between the specific $SRB_i$ curve of each activity in relation to that of the planned project ($SRB_0$). Finally, we calculated the percentage, represented by each area.

In Fig. 7 we present the different sensitivity metrics used in the SRA analysis: CI, CrI, SI, SSI. We also include the MOI and our proposal: the ARI. Although the scale is not comparable between different metrics, the ARI is the only one to offer a normalized result: if we add its value for each activity, the total sum is 100%.

With the previous results, we see how the order of priority differs according to the chosen indicator. The value obtained for each activity in all the chosen indicators is not so important because they have different meanings, rather the difference between the activities in each type of indicator. We stress that the ARI metric obtained for each activity is a normalized values and that the sum of all of them is the unit (or 100%).

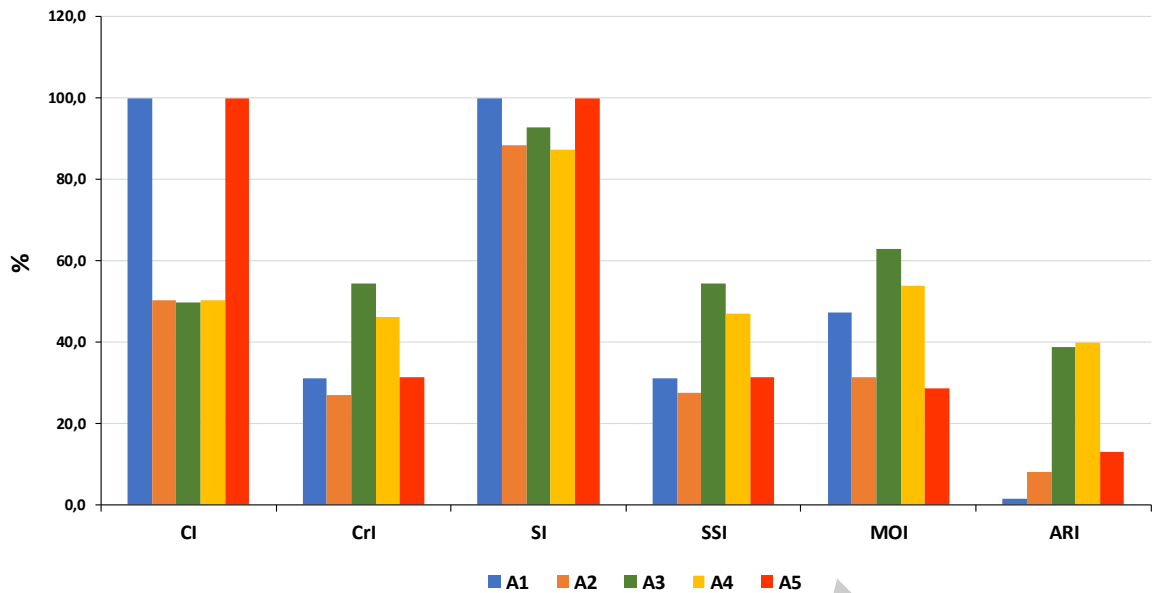

**Fig. 7** Sensitivity analysis of activities by applying different prioritization methodologies

The latter helps us to prioritize the most important activities in the project and those that need to be paid less attention to.

In Table 3 we represent the order of importance of this simple project according to the indicator chosen for its prioritization.

| Priority Order | CI | CrI | SI | SSI | MOI | ARI |
|---|---|---|---|---|---|---|
| 1 | A1 -A5 | A3 | A1 - A5 | A3 | A3 | A4 |
| 2 |  | A4 |  | A4 | A4 | A3 |
| 3 | A2- A4 | A5 | A3 | A5 | A1 | A5 |
| 4 |  | A1 | A3 | A1 | A2 | A2 |
| 5 | A3 | A2 | A4 | A2 | A5 | A1 |

**Table 3** Prioritization of activities by the indicator

If we use the CI indicator, activities A1 and A5 are the most important with 100% importance because this indicator measures the probability of these activities being on the critical path. In the example project, the two activities will always be on the critical path. On the other hand, the importance of the other activities is similar, about 50%.

In contrast, the CrI indicator prioritizes activity A3 first. This indicator analyses the correlation between the activity's duration and project duration. We observe that this activity is that which confers the most uncertainty and, consequently, it has the strongest influence on total project duration. According to this indicator, special attention should be paid to A3 because it poses the highest risk for the project.

According to Vanhoucke (2010), the SSI indicator provides more complete realistic information on the sensitivity of activities. It includes information about probability (through the CI indicator) and impact by using activity and project variability for calculations. From the results, we see that activity A3 is the most important, followed by activity A4. The MOI indicator offers similar results to previous ones. In this case, the difference with the previous indicator is because the latter incorporates the network structure as a variable, which is programmed using the successor activities of each activity.

Finally, the ARI indicator informs us that the most important activity is A4, and is almost as important as A3. In this case, these two activities confer the project more uncertainty. However, this indicator takes into account the network structure by attaching more importance (paying more attention) to the activities that are later executed.

We check how, depending on the chosen indicator, some activities are more important than others and, by selecting a different methodology, how prioritization does not coincide with the previous result. It is true that, as mentioned in the previous section regarding the different existing metrics, the deficiencies that they present are known, and offer different metrics and results so that no absolute consensus is reached to determine the most appropriate indicator.

## 6. Conclusions

Project control is an essential activity to achieve project objectives. This control can be carried out at either the project or activity level. If control is carried out at the project level, we can use tools and indicators based on the Earned Value Methodology (EV, ES, SPI, SPI(t),...) and the Earned Duration Methodology (TED, ED, DPI, EAC(t),...). We can even employ indicators that incorporate uncertainty, such as the SCoI/CCoI Methodology (Pajares and López-Paredes, 2011) or the Triad Methodology (Acebes et al., 2014b).

In this case, control is performed at the activity level and one of the most widely used simulation techniques for project control is the SRA. With this technique, and with any of its procedures, the intention is to find the most important project activities that require more attention because they could pose a risk for project objectives.

Even though there are many indicators or metrics that allow the prioritization of activities (CI, Cri, SI,...), none takes into account the project's "total risk". Here "total risk" is understood as the accumulated value of the uncertainty of project duration from the time the project starts being executed until it ends, which we refer to as the Schedule Risk Value (SRV). In other words, our metric uses the Risk Baseline instead of the cost or schedule baselines.

The ARI that we herein propose provides us with information about those activities that contribute the most to project uncertainty on the whole. We explain how an activity confers the project uncertainty from the time it begins and will maintain its uncertainty level until the activity begins. The ARI metric allows us to carry out the qualitative analysis of these activities so as to pay them the necessary attention to benefit the project as a whole.

To calculate this indicator, the most important criteria regarding activities are taken into account, such as their variability (risk) and also the network structure, by attaching more or less importance to the activity depending on its location in the network. However, what characterizes this indicator is that it is calculated using the project's risk evolution from the beginning to the end of the project. Thus it does not focus solely on the project's uncertainty value at the initial time, but on its entire evolution.

We are used to evaluating each individual project risk mainly with qualitative analyses by employing probability: impact matrices and assigning a value to each risk. With the proposed indicator, apart from prioritizing activities based on the variability (risk) they contribute to the project, we quantify that magnitude. This allows us to determine what type of strategy we should implement against risks: minimize probability or impact, eliminate the risk or transfer it. The ARI is a metric that facilitates the prioritization of activities in "probability x impact" terms. We use Monte Carlo simulation to measure the "probability" factor, and the SRB to measure the "impact" factor given the statistical properties and the position in the network of each activity.

The indicator is herein proposed to prioritize project activities using an educational project as an example. Studying the configuration of the network and its influence on the value that the indicator can take for each activity could be an interesting research line. For this purpose, the magnitude of the ARI could be related to the indicator of the series/parallel network structure, or other similar indicators, related to the project network configuration.